\documentclass[twocolumn]{aastex63}

\shorttitle{\textsc{X-ray Main Sequence}}
\shortauthors{Rodriguez}
\graphicspath{{./}{figures/}}

\usepackage{float}
\usepackage{appendix}
\usepackage{xcolor}
\usepackage{amsmath}

\begin{document}

\title{From Active Stars to Black Holes: A Tool for the SRG/eROSITA X-ray Survey and New Discoveries as Proof of Concept}

\correspondingauthor{Antonio C. Rodriguez}
\email{acrodrig@caltech.edu}

\author{Antonio C. Rodriguez}

\affiliation{California Institute of Technology, Department of Astronomy \\
1200 East California Blvd\\
Pasadena, CA, 91125, USA}

\begin{abstract}
Galactic X-ray sources are diverse, ranging from active M dwarfs to compact object binaries, and everything in between. The X-ray landscape of today is rich, with point source catalogs such as those from \textit{XMM-Newton}, \textit{Chandra}, and \textit{Swift}, each with $\gtrsim10^5$ sources and growing. Furthermore, X-ray astronomy is on the verge of being transformed through data releases from the all-sky SRG/eROSITA survey. Many X-ray sources can be associated with an optical counterpart, which in the era of \textit{Gaia}, can be determined to be Galactic or extragalactic through parallax and proper motion information. Here, I present a simple diagram --- the ``X-ray Main Sequence", which distinguishes between compact objects and active stars based on their optical color and X-ray-to-optical flux ratio ($F_X/F_\textrm{opt}$). As a proof of concept, I present optical spectroscopy of six exotic accreting WDs discovered using the X-ray Main Sequence as applied to the \textit{XMM-Newton} catalog. Looking ahead to surveys of the near future, I additionally present SDSS-V optical spectroscopy of new systems discovered using the X-ray Main Sequence as applied to the SRG/eROSITA eFEDS catalog. 
\end{abstract}

\section{Introduction}
The discovery and characterization of Galactic X-ray sources in the last sixty years opened a new window to the sky and created the field of high energy astrophysics. X-ray observations of M dwarfs, the most common stars in the Milky Way and most common hosts to exoplanets, revealed that these stars are commonly coronally active \citep[e.g. X-rays were detected from 87\% of M and K dwarfs within 7 pc of the Sun by][]{1995schmitt}. X-ray astronomy has also led to the discovery of all of the known accreting neutron star\footnote{Radio astronomy has, of course, contributed significantly to our knowledge of neutron stars as well.} (NS) systems and nearly all stellar-mass black hole (BH) systems \citep{2023tauris}, informing our understanding of the most extreme physical environments that cannot be replicated on Earth.

However, it often takes multiwavelength information to decipher the true nature of X-ray sources. Famously, the first stellar-mass black hole in the Milky Way, Cygnus X-1, was initially discovered through an X-ray sounding rocket in 1964 \citep{1965cyg}. However, it remained uncharacterized for over half a decade until an optical (and radio) source was associated with the X-ray position in 1971 \citep{1971cygOpt, 1971cygRad}. Precise radial velocity measurements of the optically bright star at the position of Cygnus X-1 were the only way to securely establish that it was a binary companion to a stellar-mass black hole \citep{1972rv, 1972bolton}. 

It is with the goal of combining X-ray + optical data to characterize Galactic astrophysical sources that I write this paper. In the days of Cygnus X-1, large error boxes associated with X-ray sources made it difficult to make multiwavelength associations. The pioneering Roentgensatellit X-ray mission \citep[ROSAT;][]{1982rosat1, 1999rosat2, 2016rosat3} was the first to image the entire sky with sub-arcminute resolution\footnote{Though the High Energy Astrophysical Observatory-2 (HEAO-2), also known as the \citep[\textit{Einstein Observatory;}][]{1979einstein} was the first to image the X-ray sky with 2" resolution, it did not conduct an all-sky survey.}, and discovered nearly 135,000 X-ray point sources with error circles of radii $\lesssim 40$" \citep[e.g.][]{2009agueros}. However, association with optical sources at that resolution is still difficult, and X-ray + optical association of ROSAT sources in the Galactic plane is nearly impossible \citep[e.g.][]{2018salvato}. 

Today, the landscape of X-ray astronomy is very different. Point source catalogs from X-ray missions that have been active for over 20 years each contain $\gtrsim 10^5$ point sources and have $\sim$few arcsecond localizations ($\theta$): \textit{XMM-Newton} 4XMM-DR13 Catalogue \citep[600,000 sources, $\theta\sim3"$;][]{2020xmm}, Second \textit{Chandra} Source Catalog \citep[350,000 sources, $\theta\sim2"$;][]{2018chandra2}, \textit{Swift}/XRT Point Source Catalog \citep[300,000 sources, $\theta\sim5"$;][]{2020swift}. X-ray astronomy is on the verge of being transformed through data releases from the all-sky SRG/eROSITA survey, with millions of X-ray sources localized to a few arcseconds \citep{2021erosita, 2021sunyaev}. 

The landscape of optical astronomy is also very different, with precise astrometry from \textit{Gaia} enabling the distinction between Galactic and extra sources \citep{2016gaia}. Large-scale time-domain photometric surveys such as the Zwicky Transient Facility \citep[ZTF;][]{bellm2019}, Transiting Exoplanet Survey Satellite \citep[TESS;][]{2015tess}, All Sky Automated Survey for Supernovae \citep[ASAS-SN;][]{2017asas-sn}, and Asteroid Terrestrial-impact Last Alert System \citep[ATLAS;][]{2018atlas} provide variability information and add a new dimension to optical datasets. Optical astronomy is also on the verge of transformation thanks to the Rubin Observatory Legacy Survey of Space and Time (LSST), which will obtain photometry for nearly an order of magnitude more sources than current time-domain optical surveys \citep[e.g.][]{2019lsst}. Just as revolutionary are the  multiplexed optical spectroscopic surveys. Millions of spectra are already in the catalog of the Sloan Digital Sky Survey \citep[SDSS;][]{2000sloan}, and SDSS-V is rapidly increasing that number as well as bringing the advent of multiplex time-domain spectroscopy \citep{2017sdssv}. Other surveys include the  Dark Energy Spectroscopic Instrument \citep[DESI;][]{2016desi}, the 4-metre Multi-Object Spectrograph Telescope \citep[4MOST;][]{20194most}, and the William Herschel Telescope Enhanced Area Velocity Explorer \citep[WEAVE;][]{2012weave}.

In this paper, I crossmatch the \textit{XMM-Newton} 4XMM-DR13 point source catalog and the SRG/eROSITA eFEDS catalog with \textit{Gaia} to just select Galactic sources. I then present a tool I name the ``X-ray Main Sequence", which distinguishes between two main types of Galactic X-ray sources: accreting compact objects (containing a WD, NS, or BH) and active stars. I put forth an empirical cut to separate between the two main types of sources. Finally, I discuss the origin of this clean separation, based on the X-ray saturation properties of solar and late-type stars.

In Section \ref{sec:data}, I present cleaned versions of the \textit{XMM-Newton} and SRG/eROSITA eFEDS catalogs, and the \textit{Gaia} crossmatch. I also describe each of the Galactic X-ray source classes and the catalogs that I use to provide classifications. In Section \ref{sec:xms}, I present the X-ray Main Sequence. In Section \ref{sec:results}, I present optical spectroscopy of new sources as a proof of concept demonstration, including a crossmatch with early data from SDSS-V. Finally, in Section \ref{sec:discussion}, I provide details on how to make best use of this diagram and why such clean distinction between source classes is possible.  

\section{Data}
\label{sec:data}
\subsection{XMM-Newton Source Catalog}
I began with the Fourth \textit{XMM-Newton} Source Catalog, 13th Edition \citep[4XMM-DR13;][]{2020xmm}, which contains 656,997 sources. It is comprised of all publicly available observations taken with the European Photon Imaging Camera (EPIC) in the 0.2--12 keV range between February 13, 2000 and December 31, 2022. Taking overlapping fields into account, this catalog covers $\sim$1328 deg$^2$ (3\%) of the sky. I selected only point sources that do not have any quality flags by making the following cuts:
\begin{itemize}
    \item Well localized point sources: \texttt{SC\_EXTENT} = 0 and \texttt{CONFUSED} = 0.
    \item 5$\sigma$ detections \citep[with a threshold as described in ][]{2020xmm}: \texttt{SC\_DET\_ML} $>$ 14.
    \item Low probability of being a spurious detection: \texttt{SC\_SUM\_FLAG} $\leq$ 1.
\end{itemize}

After these cuts, 368,068 (56\% of initial sources) remained. 99\% of sources in this final sample have a positional error of (\texttt{SC\_POSERR}) $<$ 1.95".

\subsection{Gaia EDR3 Crossmatch}
I then crossmatched the cleaned sources from the \textit{XMM-Newton} catalog with \textit{Gaia} Data Release 3 \citep[DR3;][]{2021gaiaedr3, 2023gaia_dr3} within a 2" radius\footnote{The astrometry and photometry in the Early Data Release (EDR3) is identical to that of the newer DR3.}. I did not account for proper motions since only sources with exceptionally high proper motions ($\gtrsim 100$ milliarcsec) would move more than 2" in the maximum $\sim$20 year difference between \textit{XMM-Newton} and \textit{Gaia} observations. Additionally, by using the \textit{stacked} XMM catalog, it is difficult to account for proper motions for sources having been observed potentially years apart. I only kept sources for which there is a single match within 2" --- 11,161 of the 368,068 clean sources have more than one match, and are predominantly located near the Galactic Center or the centers of clusters, but are excluded here.

I then performed the following cuts on \textit{Gaia} data to ensure both good quality and that the sources are Galactic:
\begin{itemize}
    \item Significant parallax and proper motions: \texttt{parallax\_over\_error} $>$ 3 and \\ \texttt{pm/pm\_error} $>$ 5 in both RA and DEC.
    \item Significant photometry: \\ \texttt{phot\_mean\_flux\_over\_error} $>3$ in all bands.
    \item Uncontaminated photometry: \\ \texttt{phot\_bp\_rp\_excess\_factor\_corrected} $<$ 0.05.
    \item Good astrometry: \texttt{RUWE} $<$1.4.
\end{itemize}

This cut left 25,050 sources (3.8\% of the original 4XMM-DR13 catalog). I note that I employed rather conservative cuts, particularly on the parallax. Many Galactic sources could have significant proper motions, while not a parallax. However, the goal of this work is to create a catalog of high confidence Galactic sources.

% \subsection{Extinction Correction}

% At this point, one can create either an \textit{observed} diagram using these sources, or an \textit{extinction corrected} diagram, which requires the additional assumption of a Galactic extinction model. I adopt the 3D extinction map of \cite{2019green}, but ultimately present both \textit{observed} and \textit{extinction corrected} diagrams. I choose this map since it takes advantage of well-measured distances from \textit{Gaia}, and extends out to a few kiloparsecs, as opposed to all-sky maps which only extend out to 1 kpc \citep[e.g.][]{2023dustmap_south}. Since the map of \cite{2019green} is based on Pan-STARRS-1 data, I only keep sources with DEC $>-30^\circ$. This final cut yields 4,493 sources. 

\subsection{Object Classes and Their Catalogs}
\label{sec:classes}
I used a variety of modern catalogs to provide classifications for known Galactic X-ray objects. As none of these catalogs are by any means complete\footnote{For example, I used the classic Ritter and Kolb catalog of cataclysmic variables (CVs), which contains some of the most well known, spectroscopically verified CVs. This catalog contains 1,429 sources, while the larger Open CV Catalog \citep{2020opencv}, contains over 10,000 sources (mostly candidates).}, I instead focused on a high purity fraction for each source classification. I crossmatched all catalogs with the XMM-Newton/\textit{Gaia} catalog described above to obtain consistent X-ray and optical fluxes. X-ray binaries, symbiotics, and spider binaries are exceptions, where the number of sources is low enough that I used X-ray detections from any available detection. 

\subsubsection{Active Single Stars}
Active stars are soft X-ray emitters due to their rapid rotation, which leads to coronal activity \citep[e.g.][]{1981pallavicini}. Due to their fully convective nature and overall abundance, active M dwarfs are the most common X-ray emitting single stars \citep{2013m_dwarf}\footnote{Other esoteric classifications also exist: BY Dra stars are K and M dwarfs which show photometric variability on their rotation period \citep{1983by_dra}, and FK Com stars are G and K giants which show similar variability \citep{1981fk_com}. Both subtypes are rapid rotators, some of which show evidence for binarity, though the binary fraction of each subtype is unknown.}. I used the sample of 823 active stars from \cite{2011wright}. These systems are located both in the field and in open clusters, and have X-ray detections originally from the ROSAT Bright Source Catalog \citep{1999voges}. After crossmatching with my XMM-Newton/\textit{Gaia} catalog, I was left with 112 sources.

\subsubsection{Active Binary Stars: RS CVn Systems}
RS CVn stars are binaries typically consisting of a slightly evolved subgiant and a solar or late-type star, which may or may not be evolved \citep{2003rs_cvn}. Their orbital periods range from 5--6 hours to tens of days, inducing fast rotation of the stars. This leads to increased activity and X-ray emission \citep{1981rscvn}. RS CVns have garnered some attention as being contributors to the Galactic Ridge X-ray Excess \citep{1983worrall}, though the contribution of cataclysmic variables is comparable or even higher \citep{2006revnivtsev}.

There are a handful of well-studied RS CVn systems that receive much attention in the literature \citep[e.g.][]{2021rscvn_example}, but I encountered a dearth of large catalogs of vetted systems. Instead, I assembled a list of objects from the International Variable Star Index (VSX) in the \texttt{RS} category, which yields 73,320 sources. After crossmatching with my XMM-Newton/\textit{Gaia} data set, this left only 341 systems. Since various well-known RS CVn systems are catalogued as eclipsing binaries (\texttt{EB*}) in the Simbad database, I also included as active binaries systems from my XMM-Newton/\textit{Gaia} crossmatch that are labeled as \texttt{EB*} and have $N>5$ references in Simbad (147 systems). It appears as though the time is right for a dedicated survey of X-ray emitting RS CVn systems.

\subsubsection{Young Stellar Objects (YSOs)}

YSOs are dynamic environments, with the seeds of planets forming in the circumstellar disk extending to hundreds of stellar radii \citep[e.g.][]{2016ysoreview}. The pre-main sequence star at the center is inflated, highly magnetic, and is born rotating rapidly \citep[e.g.][]{1986rotationyso}. This high rotation makes YSOs perfect candidates for being X-ray sources, which were first discovered to be as such by \cite{1981feigelson}. However, the high levels of X-ray luminosity, hot (0.5--5 keV) temperatures, variability on hour timescales pointed to a sources of X-rays in addition to coronal activity, such as magnetic reconnection \citep{2002feigelson}.

I used the catalog of YSO candidates assembled using a crossmatch of \textit{Gaia} and \textit{WISE} mid-infrared data \citep{2019yso_gaia}. That catalog was assembled using a training set of well-vetted YSO catalogs to which a Random Forest classifier was used to infer probability of an object being a YSO. I began with all objects in that catalog that have a 96\% probability or greater of being a YSO. Of those, 398 are in my XMM-Newton/\textit{Gaia} crossmatch.

\subsubsection{Cataclysmic Variables}
Cataclysmic variables (CVs) are close binaries where a WD accretes from (typically) a late-type main sequence companion \citep[e.g.][]{1995warner}. In (non-magnetic) CVs where the accretion disk extends down to the WD surface, X-rays originate from the disk-WD boundary layer, while in magnetic CVs where the field is strong enough ($B\gtrsim$ 1 MG) to influence the accretion, X-rays originate from the accretion shock on the WD surface \citep{2017mukai}. CVs are particularly interesting X-ray sources since they are thought to be the dominant contributors to the excess of X-rays from the Galactic Ridge and Galactic Center \citep{2006revnivtsev, 2016hailey}. 

AM CVn systems are ultracompact CVs ($P_\textrm{orb}\approx 5-65$ min), where a WD accretes from a helium-dominated degenerate or semi-degenerate companion. While less than 100 of these systems are known, they are particularly interesting in that some of these will be among the loudest sources of gravitational waves as seen by the Laser Interferometer Space Antenna \citep[e.g.][]{2001nelemans, 2017lisa}

I used the most updated publicly available catalog of cataclysmic variables (CVs) to date, the Final Version (December 31, 2015) of the Ritter and Kolb catalog \citep{2003ritter}. This catalog contains 1,429 systems primarily discovered through their optical outbursts and/or X-ray associations. 64 of those systems are present in my XMM-Newton/\textit{Gaia} catalog, including magnetic CVs. Intermediate polars (DQ Her stars) and polars (AM Her stars) are labeled as subtypes \texttt{DQ} and  \texttt{AM}, respectively, and AM CVn systems are labeled as type \texttt{AC}.

I also included supersoft X-ray sources (SSSs), which are WDs that have a layer of steadily burning hydrogen on their surface. Their X-ray spectra have equivalent blackbody temperatures ranging from 15--80 eV, which means that little of their bolometric flux overlaps with the \textit{XMM-Newton} energy range. I used the catalog of SSSs from \cite{1997sss} and kept only those systems which have a detection in the \textit{XMM-Newton} source catalog. 4 systems fulfill this condition (one is AG Dra, which is a symbiotic SSS and labled as ``symbiotic" in Figure \ref{fig:xms}).

\subsubsection{Millisecond Pulsar (Spider) Binaries}
Millisecond pulsars have been found in close binaries with either an M dwarf or brown dwarf companion --- redbacks and black widows, respectively. These systems are X-ray sources due to the presence of an intrabinary shock which converts the pulsar power to X-rays \citep[e.g.][]{2016romani}. While these systems are not accreting, for the purposes of this paper, I will include them in under the umbrella of compact objects, as these systems occupy the upper left portion of the X-ray main sequence. I took the catalog from \cite{2023spiders}, which compiles both \textit{Gaia} positions of known redbacks and black widows as well as their X-ray detections in the literature (all detections come from instruments that would lead to at most a factor of 2--3 correction to the flux in the 0.2--12 keV range, so I kept original measurements). I omitted all sources labeled as ``candidates", and kept only sources that had an X-ray detection, which yielded 15 redbacks and 4 black widows.

\subsubsection{Neutron Star and Black Hole X-ray Binaries: Low-mass, High-mass, and Ultracompact}

X-ray binaries (XRBs) are systems in which a neutron star or a black hole accretes from a binary companion. This takes place in the form of Roche lobe overflow from a degenerate donor star in ultracompact XRBs (UCXBs), Roche lobe overflow from a late-type donor star in low-mass XRBs (LMXBs), and wind accretion from an O/B type star in high-mass XRBs (HMXBs) \citep[e.g.][]{2023tauris}. Most of these systems have been discovered through outburst events, which lead to a transient brightening in both X-rays and optical luminosity \citep[e.g.][]{1993lewin}. A handful of systems are in a persistent ``high" state, accreting near the Eddington limit, and have been visible as bright sources since the early days of X-ray astronomy \citep[e.g.][]{1988vanparadijs}. It is not useful for me to plot LMXBs in outburst, since it defeats the purpose of creating this diagram --- to identify \textit{persistent} Galactic X-ray sources. Instead, I just plotted XRBs which are either in quiescence or in a persistent high state.

I assembled a list of quiescent or high state LMXBs and HMXBs from three papers: \cite{1999menou} has archival X-ray fluxes of NS and BH binaries before \textit{Chandra} and 
\textit{XMM-Newton}, \cite{2001garcia} reports results primarily of quiescent BH LMXBs from early \textit{Chandra} data, and \cite{2006russell} assembles all known XRB X-ray measurements, distinguishing between quiescent and outbursting sources, neutron stars and black holes. For the UCXBs, I use thed catalog from \cite{2023ucxb}, which exclusively has NS accretors.

In all cases, I only kept sources that 1) have a significant \textit{Gaia} detection and flux measurement (i.e. bright enough and not located in a globular cluster), 2) have detections in a well-defined persistent high state or quiescent state, and 3) have a well-measured X-ray flux  (i.e. not an upper limit or marginal detection). This left 2 UCXBs, 4 HMXBs, and 9 LMXBs. 

\subsubsection{Symbiotic Stars and Symbiotic X-ray Binaries}

Symbiotic stars are long-period ($P_\textrm{orb}\gtrsim 200$ days) binary systems in which a NS or WD accretes from an evolved, typically a red giant, companion\footnote{Symbiotic stars with a BH accretor should in principle exist, but no such systems have been confirmed to date.} \citep[e.g.][]{2023belloni}. The term ``symbiotic X-ray binary" is used often in the literature to refer to the subset of systems in which a NS is the accretor. Such systems are prone to X-ray bursts similar to LMXBs and HMXBs. 

I used the catalog of WD accretors from \cite{2013luna} and the catalog NS accretors from \cite{2019yungelson}. To obtain X-ray detections of a few more systems, I crossmatched my XMM-Newton/\textit{Gaia} catalog with the catalog of symbiotic stars from \cite{2019symbiotic}. I adopted the same quality cuts as in the previous subsection, rendering a total of 11 symbiotic stars with a WD accretor and 4 symbiotic stars with a NS accretor.

\subsection{SRG/eROSITA eFEDS Catalog}
While data releases from SRG/eROSITA are imminent at the time of writing, the eROSITA Final Equatorial Depth Survey (eFEDS) catalog provides a preliminary look at the final projected sensitivity of the mission. A detailed summary of the survey is outlined in \cite{2022brunner}, and the counterpart association in \cite{salvato2021}. The limiting flux of eFEDS is comparable to that of the \textit{XMM-Newton} catalog, $F_X \sim 6.5\times 10^{-15} \textrm{ erg s}^{-1} \textrm{cm}^{-2}$, though with a slightly larger positional uncertainty of 4.7" \citep{salvato2021}. The full catlaog contains 27,369 sources. In order to select only Galactic sources, I took the catalog of \cite{salvato2021}, and kept only sources with \texttt{CTP\_Classification} as \texttt{SECURE GALACTIC} or \texttt{LIKELY GALACTIC}. I enforced the same astrometric quality cuts as in the \textit{XMM-Newton} catalog. Finally, I kept only sources with an X-ray detection likelihood \texttt{ERO\_DET\_ML} greater than 10, as suggested by \cite{salvato2021}. This left 1,385 sources (5.1\% of the original catalog).

\subsection{SDSS-V (SDSS DR18) Crossmatch}
\label{sec:sdssinfo}
SDSS recently released its 18$^\textrm{th}$ data release \citep[SDSS DR18;][]{2023sdssv}, which includes the first publicly available spectra from SDSS-V \citep{2017sdssv}. SDSS-V, among other multiplexed spectrographs (DESI, WEAVE, 4MOST), is undertaking time-domain optical spectroscopic surveys. SDSS-V is uniquely targeting SRG/eROSITA targets, and includes spectra in DR18 of 16,548 objects in the eFEDS patch of sky. This means that even objects without an X-ray counterpart are targeted \citep{2023sdssv}. I crossmatched the SDSS-V eFEDS catalog with the Galactic eFEDS sources described above, and kept only sources with \texttt{CLASS == STAR} and high signal-to-noise \texttt{SN\_MEDIAN\_ALL > 10}. This left 49 sources, 42 of which pass \textit{Gaia} quality cuts.

\section{The X-ray Main Sequence}
\label{sec:xms}
\begin{figure*}
    \centering    \includegraphics[width=\textwidth]{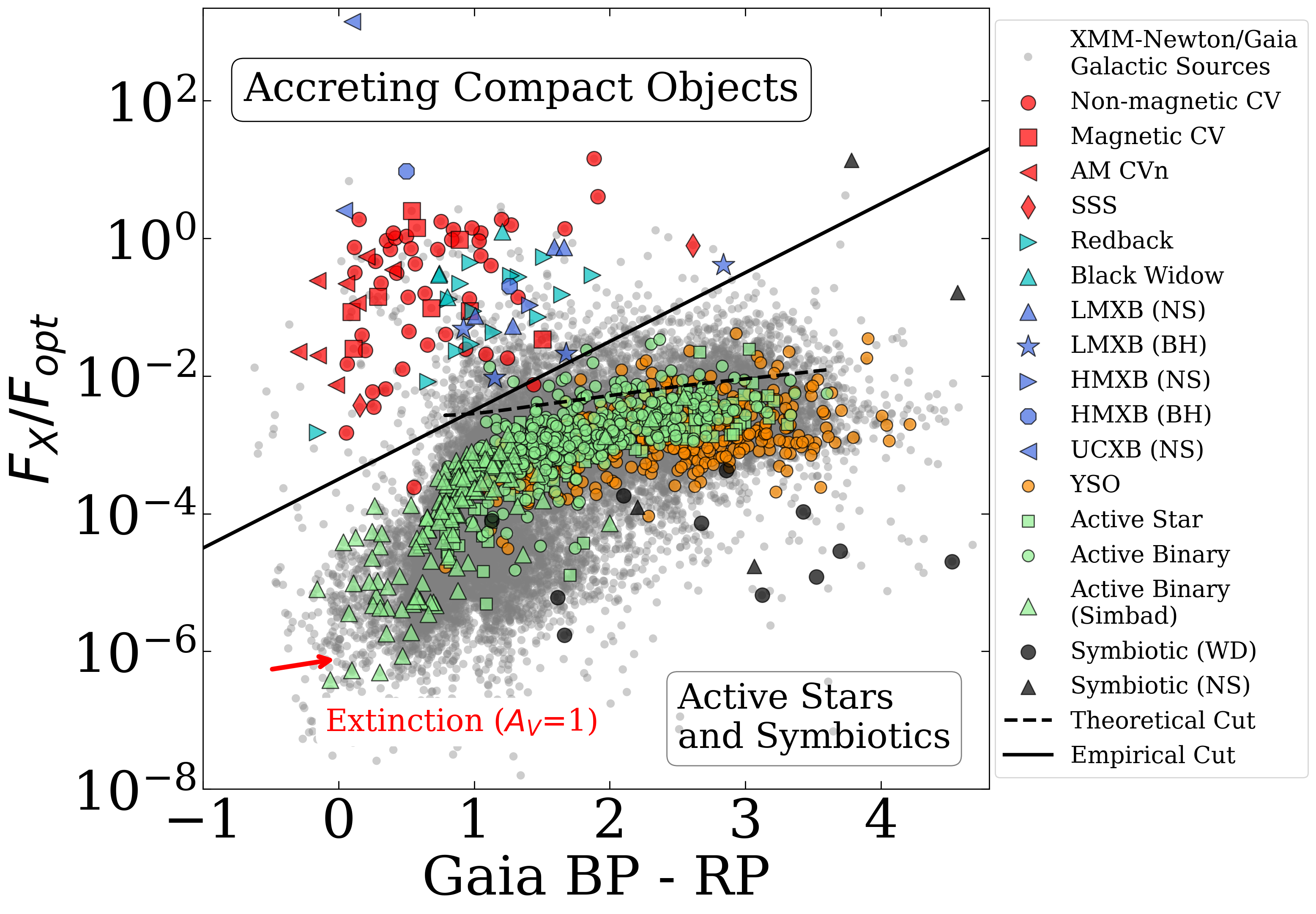}
    \caption{The X-ray Main Sequence. Galactic sources from the XMM-Newton/\textit{Gaia} crossmatch is shown in grey. Accreting compact object binaries in the upper left are separated from symbiotic and active stars on the bottom right by the ``empirical cut" (solid line) or ``theoretical cut" (dotted line). All classifications on the right side panel are from the literature, and described in Section \ref{sec:classes}. No extinction correction is applied here, but the extinction vector is shown (de-reddening slides sources towards the lower left).}
    \label{fig:xms}
\end{figure*}

In Figure \ref{fig:xms}, I present the X-ray Main Sequence with two cuts to distinguish between accreting compact objects and active stars: the ``empirical cut" and a ``theoretical cut". Symbiotic stars occupy their own region of phase space, below active stars. This means that the diagram can be used to easily separate compact objects with low-mass stellar companions from active stars, but further information is needed to separate out symbiotics (compact objects with evolved donors). The ``X-ray Main Sequence" name is attractive since main sequence stars follow their own track, while compact objects occupy separate regions of phase space\footnote{Though really, this is a color-color diagram.}.

The empirical cut is drawn by eye (with the functional form $\log_{10} y=x-3.5$, where $y$ is the vertical axis and $x$ the horizontal axis). This cut encompasses the majority of accreting compact objects, while removing active stars. More detailed analysis of where to place this cut will be presented in future work. Later in the paper, I discuss the construction of the theoretical cut based on the saturation limit of X-rays stemming from coronal activity of solar and late-type stars.  

YSOs and active stars reside below the empirical cut, with YSOs occupying the reddest regions of the diagram and active (binary) stars dominating the bluest regions. Most of the unclassified sources in the diagram have very low $F_X/F_\textrm{opt}$ values, potentially being active binaries that have not been classified as such. 

\begin{figure*}
    \centering
    \includegraphics[width=0.44\textwidth]{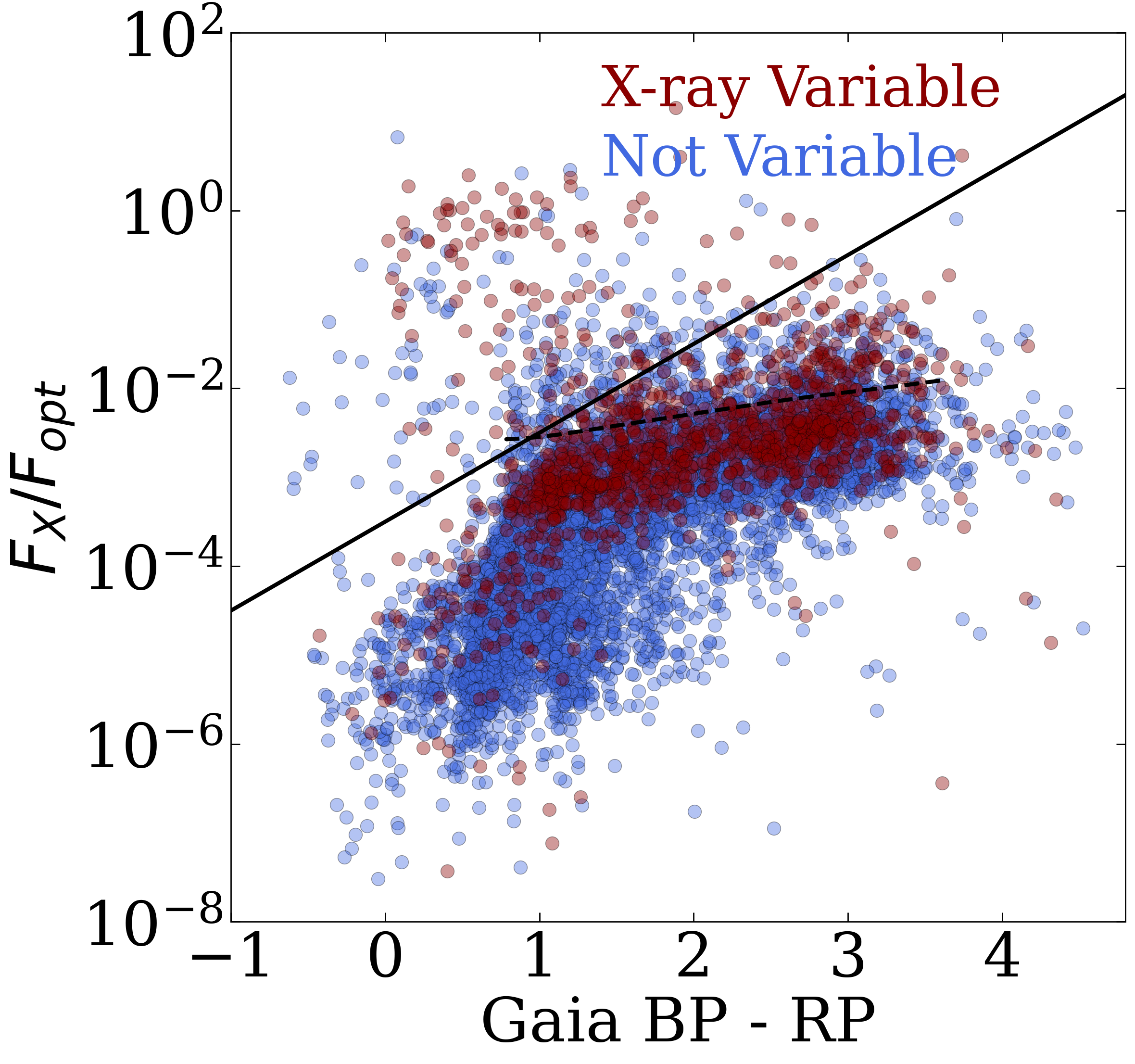}\includegraphics[width=0.53\textwidth]{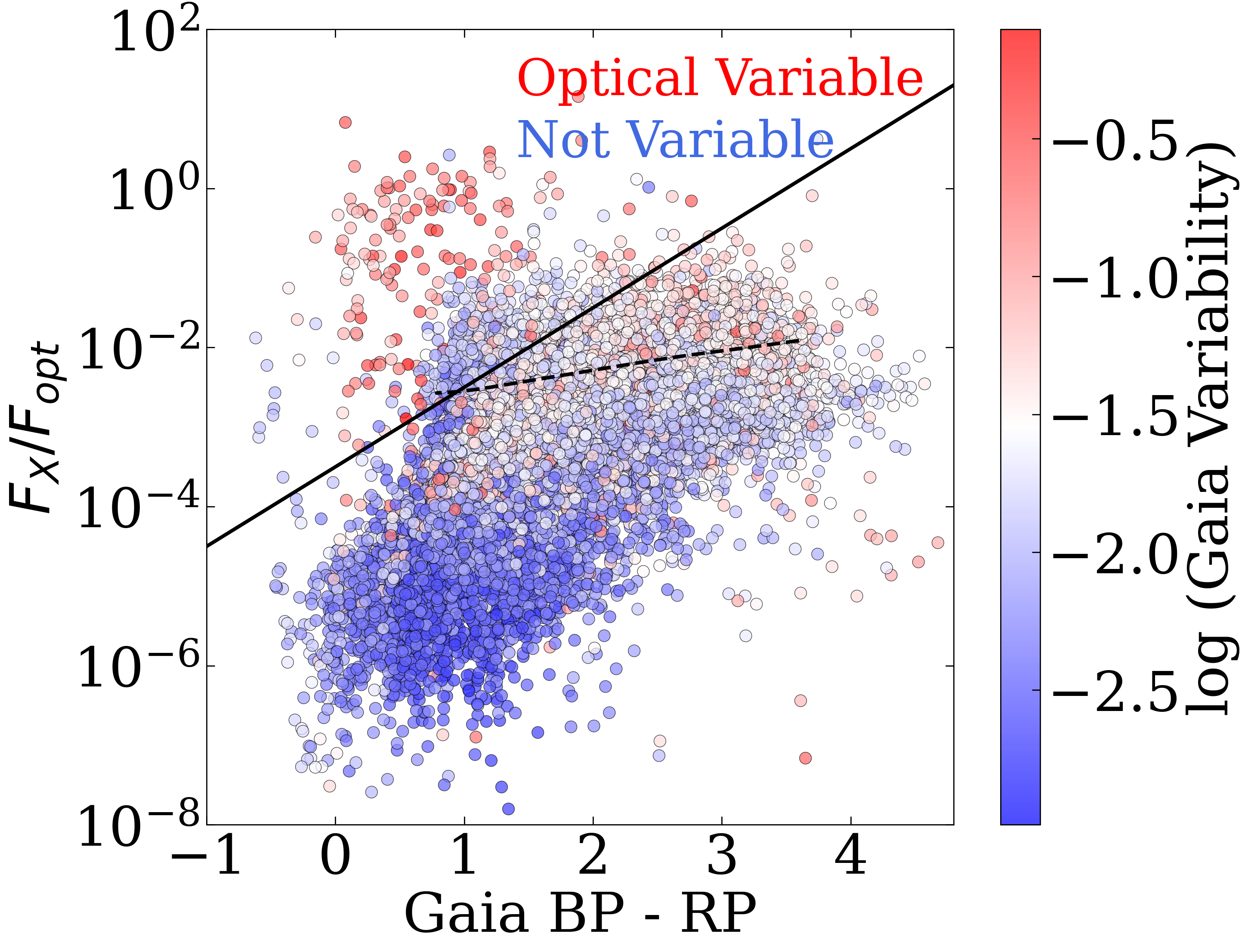}
    \caption{The same dataset as in Figure \ref{fig:xms}, but color coded by X-ray variability (\textit{left}) and optical variability (\textit{right}). In both cases, the most variable sources tend to be located above the cut or just below it.}
    \label{fig:xms_variable}
\end{figure*}

In Figure \ref{fig:xms_variable}, I plot the XMM-Newton/\textit{Gaia} crossmatch, color coded with variability metrics. In the left panel of Figure \ref{fig:xms_variable}, I color code by the X-ray variability flag, \texttt{SC\_VAR\_FLAG} (some sources are missing due to the lack of X-ray counts that can be used to compute the variability metric). While non-variable X-ray sources tend to be located everywhere in the diagram, variable X-ray sources are located either in the upper left accreting compact object corner, or in the active star corner, near the boundary.

\begin{figure}
    \centering
    \includegraphics[width=0.48\textwidth]{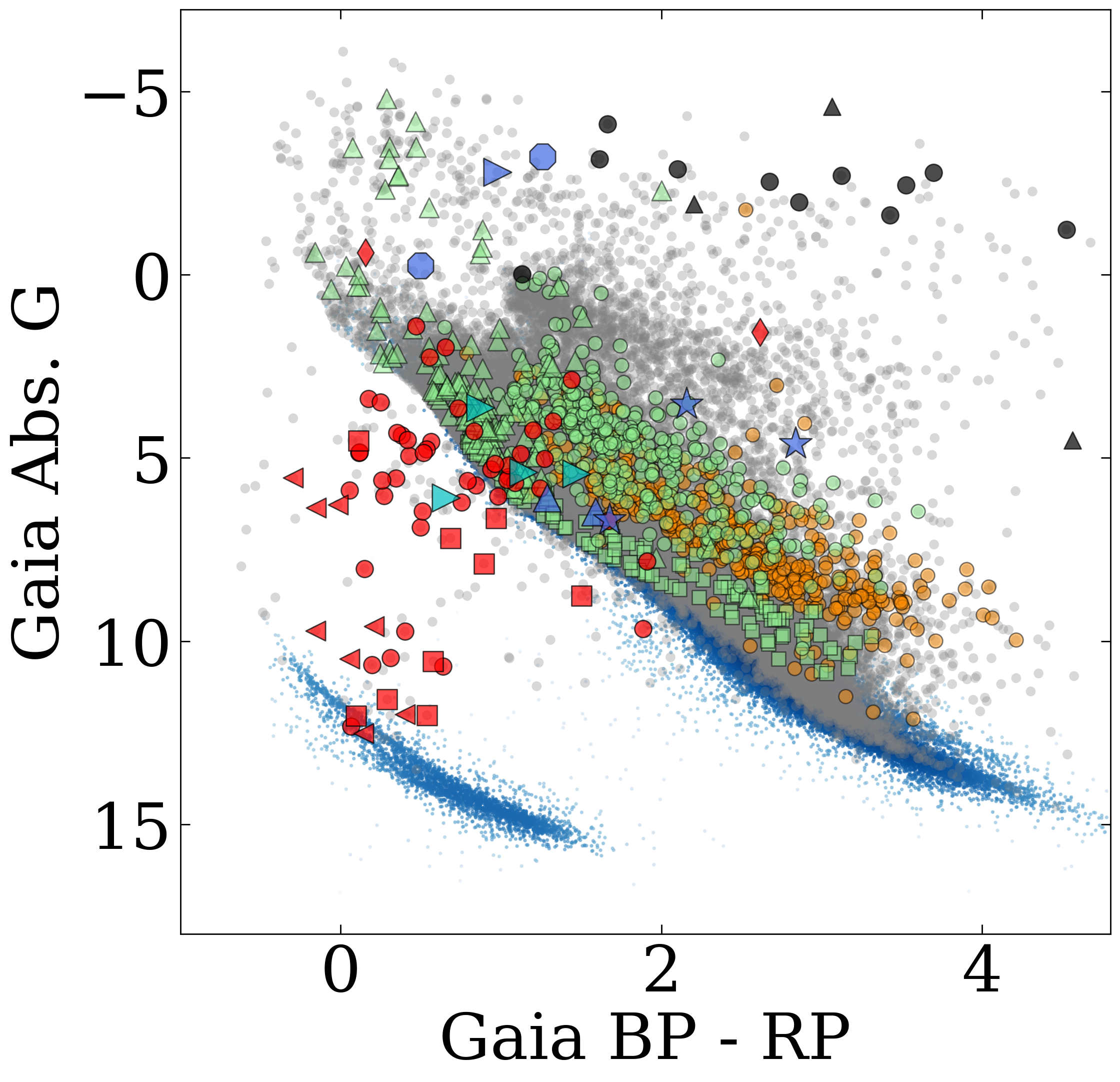}
    \caption{Objects from Figure \ref{fig:xms} (same coloring conventions) plotted atop the 100 pc \textit{Gaia} Hertzsprung-Russell (HR) diagram (light blue). The full XMM-Newton/\textit{Gaia} crossmatch is shown in gray. }
    \label{fig:xmm_hr}
\end{figure}

In the right panel of Figure \ref{fig:xms_variable}, I color code by an optical variability metric from \textit{Gaia}. This metric is essentially the mean number of standard deviations from the median flux by which the source varies: $\sigma_G \sqrt{N_\textrm{obs}}/\langle G \rangle$ \citep[e.g.][]{2021mowlavi, 2021guidry}. The most optically variable sources are, as with X-ray variable sources, either in the upper left part of the diagram or near the boundary between classes. Finally, in Figure \ref{fig:xmm_hr}, I plot all objects with a significant (3$\sigma$) parallax in the 100 pc \textit{Gaia} Hertzsprung-Russell diagram. 

\section{Results: Discovery of New Systems as Proof of Concept}
\label{sec:results}
In the following subsections, I show the application of this diagram to the discovery of several new objects. This is a pilot study which serves as a proof of concept.

\subsection{SRG/eROSITA eFEDS + SDSS-V Optical Spectroscopy}
\begin{figure}
    \centering    \includegraphics[width=0.5\textwidth]{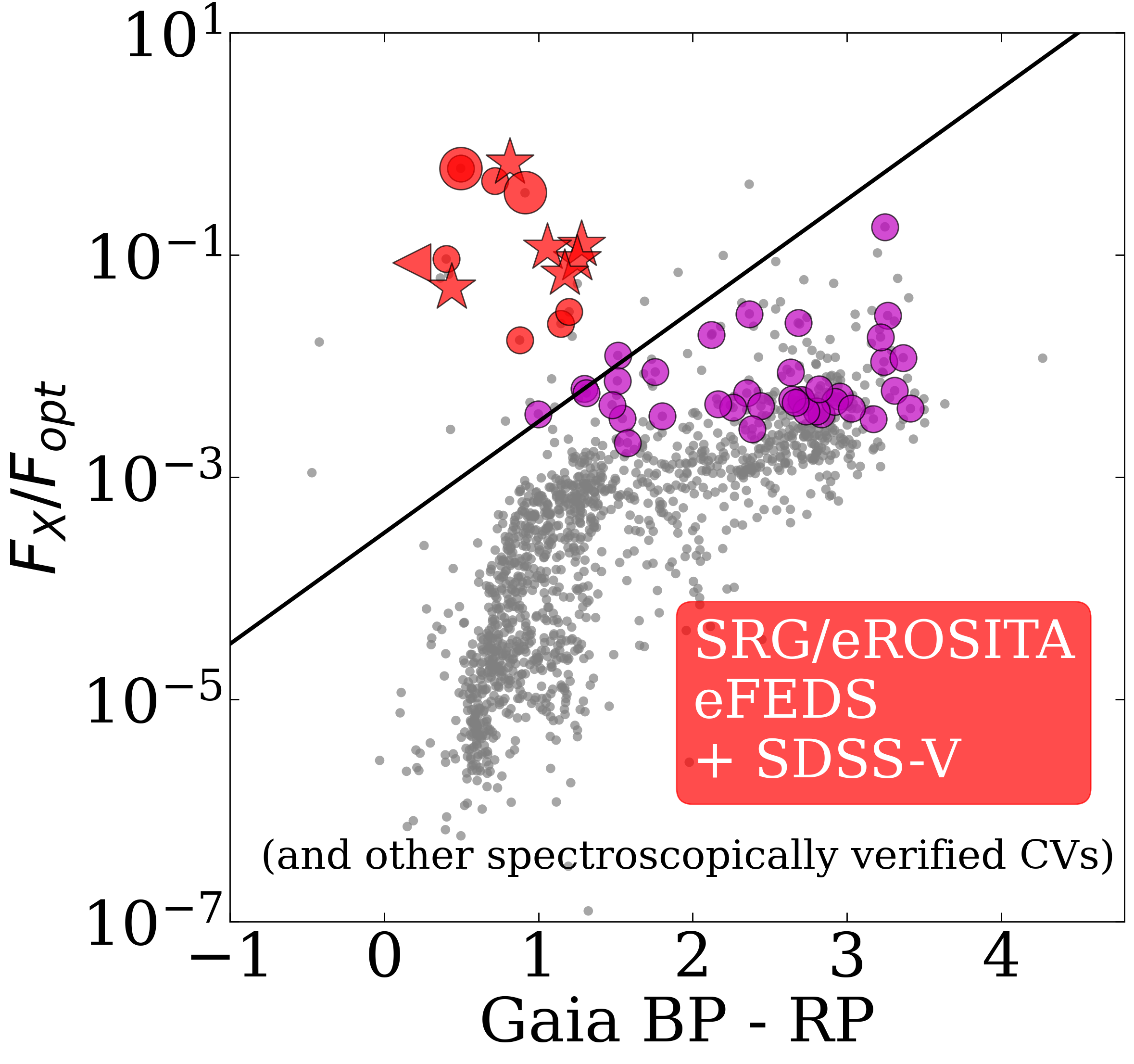}
    \caption{Galactic sources from the SRG/eROSITA eFEDS catalog (gray), with colored circles indicating those that have an SDSS-V spectrum. The two larger circles are polars from \cite{2023polar}. The same empirical cut from Figure \ref{fig:xms} distinguishes CVs (red) from active stars (magenta). Red stars are spectroscopically confirmed CVs from an ongoing survey using the \textit{XMM-Newton} catalog, and the red triangle is an AM CVn (ultracompact CV) from a separate SRG/eROSITA catalog \citep{2023amcvn}.}
    \label{fig:efeds_sdss}
\end{figure}

In Figure \ref{fig:efeds_sdss}, I present the SRG/eROSITA eFEDS catalog of Galactic objects, along with those objects that have an SDSS-V spectrum in SDSS DR18. As described in Section \ref{sec:sdssinfo}, I only keep objects with a high signal-to-noise ratio that have \texttt{CLASS == STAR}. Using the empirical cut (same as in Figure \ref{fig:xms}), I classify all objects below the cut as active stars and all above as CVs. I visually inspect all 42 spectra, and confirm the \texttt{SUBCLASS} classifications in SDSS DR18 (28 M stars, 8 K stars, 1 G star, and 5 CVs). This confirms the effectiveness of the X-ray Main Sequence, though a different empirical cut should ideally be adopted depending on the X-ray telescope, which would have a different energy range. I also note that targeted objects are predominantly located at high values of $F_X/F_\textrm{opt}$, likely due to a selection of optically faint objects (e.g. no objects in the eFEDS/SDSS-V crossmatch are brighter than $G=16$).  

\begin{figure}
    \centering
    \includegraphics[width=0.48\textwidth]{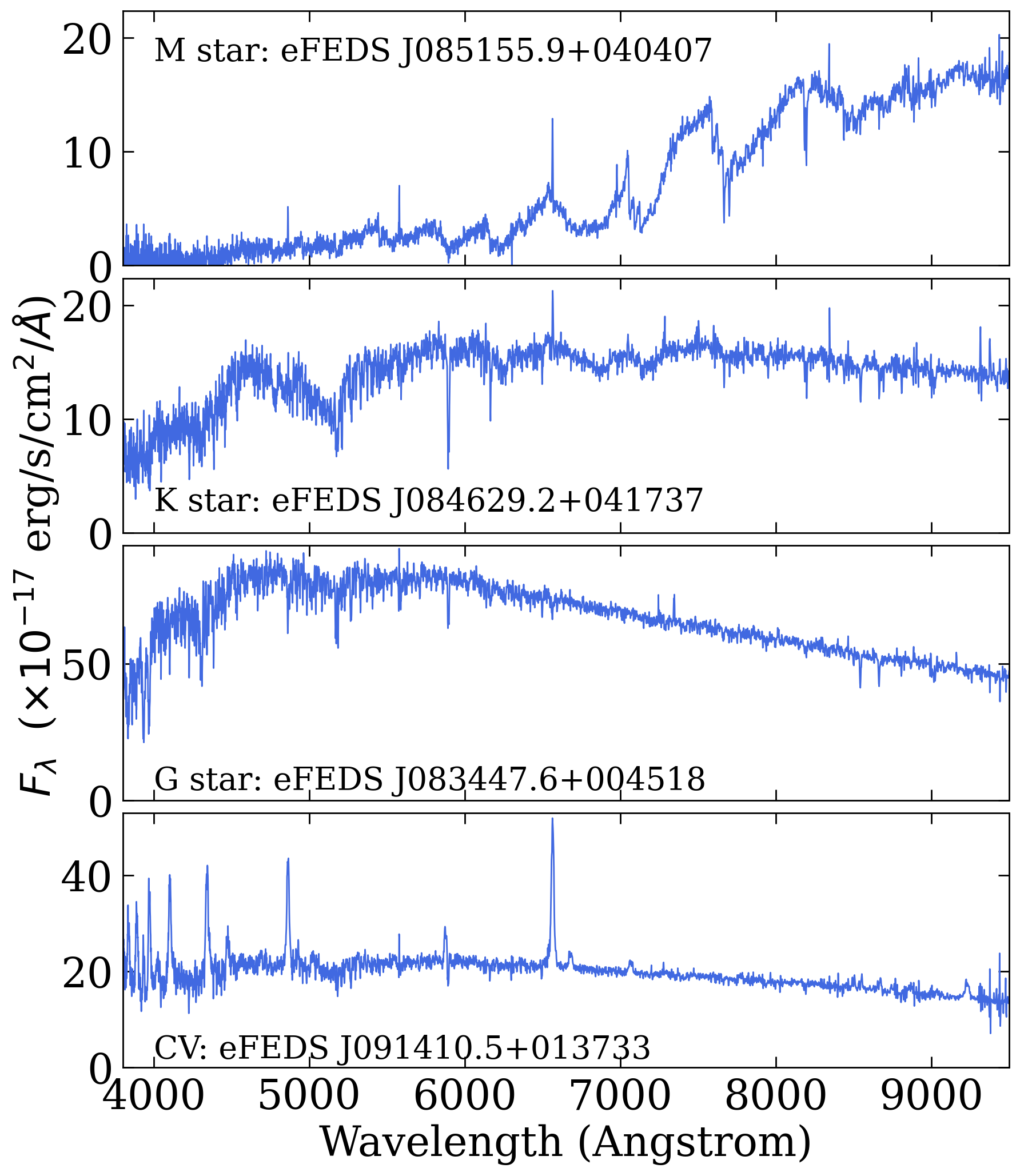}
    \caption{SDSS-V spectra of four objects with distinct spectral types in the SRG/eROSITA eFEDS catalog. All objects are new discoveries, with spectroscopy confirming the predicted classification from the X-ray Main Sequence.}
    \label{fig:efeds_sdss_spectra}
\end{figure}

The objects in the lower right part of Figure \ref{fig:efeds_sdss} are predominantly active stars, with most M and K dwarfs showing Balmer emission lines or at least H$\alpha$ in emission. However, further work is needed to determine the binary nature of these systems. In Figure \ref{fig:efeds_sdss_spectra}, I present SDSS-V co-added spectra of four objects (publicly available in DR18) in the eFEDS/SDSS-V crossmatch, one of each spectral type. All four objects are presumably new, with no references in either Simbad or VSX\footnote{At the time of writing, work is being undertaken to identify all CVs in the eFEDS field (Schwope, A. et al, in prep).}.

\subsection{Other SRG/eROSITA Systems: Magnetic CVs and AM CVns}

I have shown elsewhere that this diagram can be used to discover two important classes of CVs: magnetic CVs and ultracompact AM CVns. In \cite{2023polar}, we discovered two polars using a crossmatch of the SRG/eROSITA eFEDS catalog with \textit{Gaia} and ZTF. In \citep{2023amcvn}, we discovered an eclipsing AM CVn which showed little optical variability due to poor photometric coverage. We demonstrated that it was the X-ray + optical crossmatch and its placement in the X-ray Main Sequence that enabled the discovery and characterization of that object. I show all three systems in Figure \ref{fig:efeds_sdss}, with the two polars shown as larger red circles and the AM CVn as a triangle.

\subsection{XMM-Newton CVs with Keck and Palomar Spectroscopy}
\label{sec:xmm_cv}

To further demonstrate the effectiveness of the X-ray Main Sequence in selecting accreting compact objects, I am undertaking a spectroscopic survey of CV candidates selected from the upper left corner of the XMM-Newton/\textit{Gaia} crossmatch presented in this paper (as well as other X-ray catalogs). To complement this survey, I am also crossmatching with optical photometry from the Zwicky Transient Facility (ZTF; see Appendix \ref{sec:appendix} for details). Here, I present six systems from this ongoing survey to highlight the effectiveness of the X-ray Main Sequence. In particular, the systems I present are ``exotic CVs" --- CVs which have been particularly rare in purely optical surveys (i.e. those that search for optical outbursts). A description of of all objects and observations is in Appendix \ref{sec:appendix}. Optical spectra of all objects are shown in Appendix Figure \ref{fig:xmm_spectra}, and optical light curves in Appendix Figures \ref{fig:xmm_lc} and \ref{fig:xmm_phot}.

\section{Discussion}
\label{sec:discussion}

\subsection{How to Interpret This Diagram}
As discussed in Section \ref{sec:data}, the catalogs for all objects types in this sample are almost certainly incomplete. In no way do I claim that this plot perfectly distinguishes between classes of objects. In addition, the majority of objects in the \textit{XMM-Newton} source catalog, even those with optical counterparts, remain unidentified. So, it is difficult to obtain estimates of either purity or completeness without more detailed classification (e.g. from optical spectroscopy). The most that we can say at the present time is that the empirical cut in Figure \ref{fig:xms} introduces at most a few active stars into the accreting compact object corner, and vice versa. However, there are a few CVs and HMXBs that make their way into the bottom right corner, presumably due to high optical contribution from the donor and/or accretion disk. Despite these caveats, the diagram works very well for classifying objects actually detected by X-ray surveys. That's the point: not to find every compact object, but the classify the ones that are X-ray sources.

\subsection{Why This Diagram Works: The Saturation Line}

A plausible explanation for the clear clustering in this diagram rests on a well-known empirical result in the study of solar and late-type active stars: the X-ray luminosity ``saturates" at the (typically claimed) limit of $L_X/L_\textrm{bol} \approx 10^{-3}$. 

This existence of this X-ray saturation limit has been proposed to be due to the limit on the magnetic field strength as a result of the dynamo mechanism \citep{2022reiners}. In the earliest works investigating stellar activity, \cite{1972skumanich} connected the strength of the Ca II H and K lines with stellar rotation periods. \cite{1981pallavicini} was the first to use X-ray fluxes (from the \textit{Einstein Observatory}) as a stellar activity indicator and show a dependence of X-ray luminosity on stellar rotation periods. Later works then used larger X-ray datasets (such as ROSAT) to show that at the fastest rotation periods, X-rays saturate at $L_X/L_\textrm{bol} \approx 10^{-3}$ \citep[e.g.][]{2014reiners, 2020magaudda, 2021johnstone}. In most cases, $L_X/L_\textrm{bol}$ is plotted versus the Rossby number, \textit{Ro}, which is the ratio of the rotation period over the convective turnover timescale. There is also a ``super-saturated" regime at the lowest Rossby numbers where $L_X/L_\textrm{bol}$ turns over and decreases from the saturation limit \citep[e.g.][]{2023nunez}.

\begin{figure}
    \centering
    \includegraphics[width=0.5\textwidth]{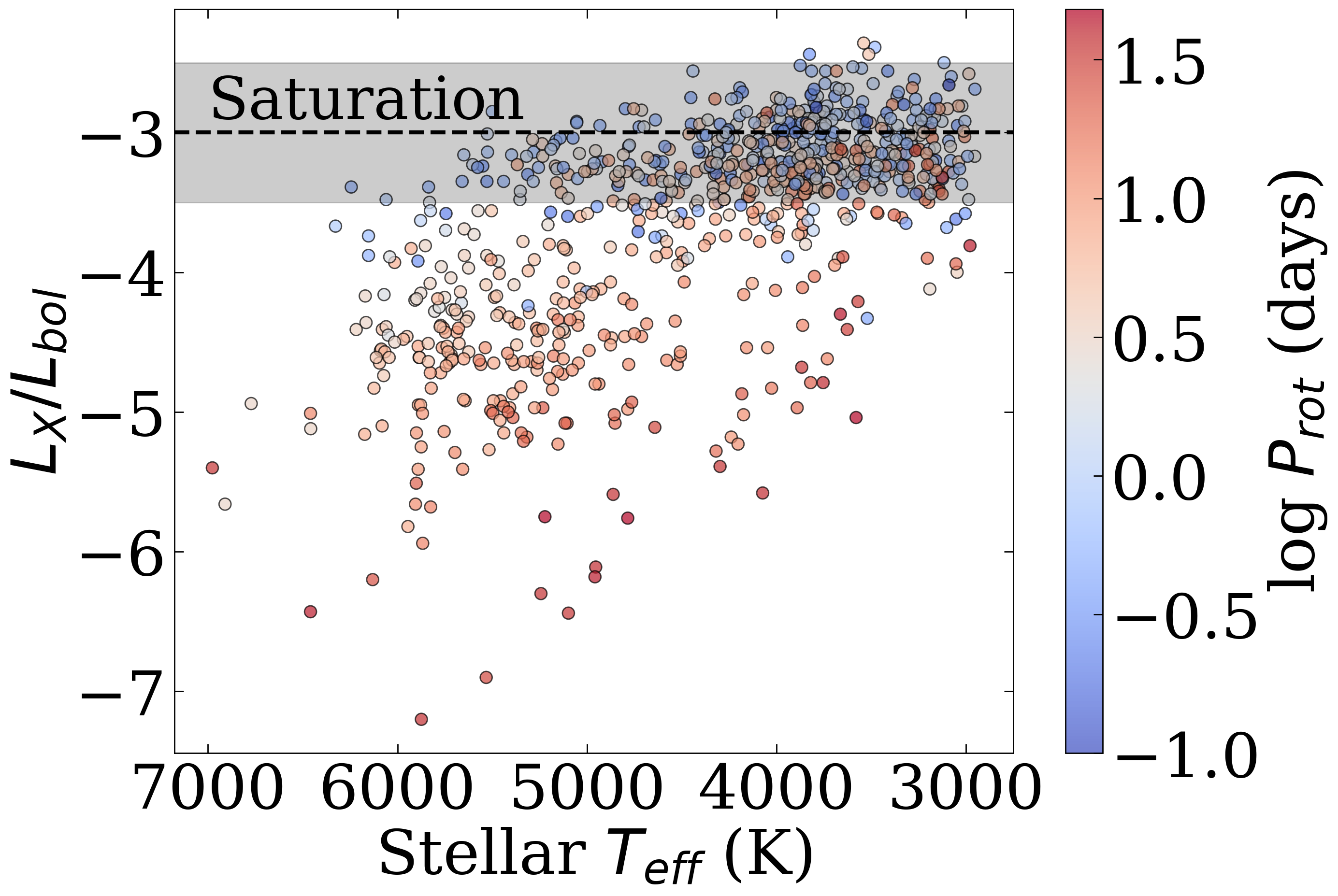}
    \caption{All active stars in the sample of \cite{2011wright} are below the saturation limit of $L_X/L_\textrm{bol} = 10^{-2.5}$. The majority of stars in the saturated regime tend to be cooler, with $T_\textrm{eff} \lesssim 5000$ K.}
    \label{fig:active}
\end{figure}

In recent years, it has been proposed that stellar activity and magnetism arise from a dynamo mechanism --- where it can be most simply stated that kinetic rotational energy is converted to magnetic energy\citep[e.g.][]{2014charbonneau}. If the dynamo theory holds, then limits on stellar rotation naturally set a limit on magnetic activity, and therefore X-ray luminosity \citep{2022reiners}. In Figure \ref{fig:active}, I present the sample of 823 active stars from \cite{2011wright}. In practice, the saturation limit has a variance of $\approx 0.5$ dex, so I adopt a saturation limit of $L_X/L_\textrm{bol} = 10^{-2.5}$ when creating the ``theoretical cut" in the X-ray Main Sequence.

\begin{figure*}
    \centering
    \includegraphics[width=0.8\textwidth]{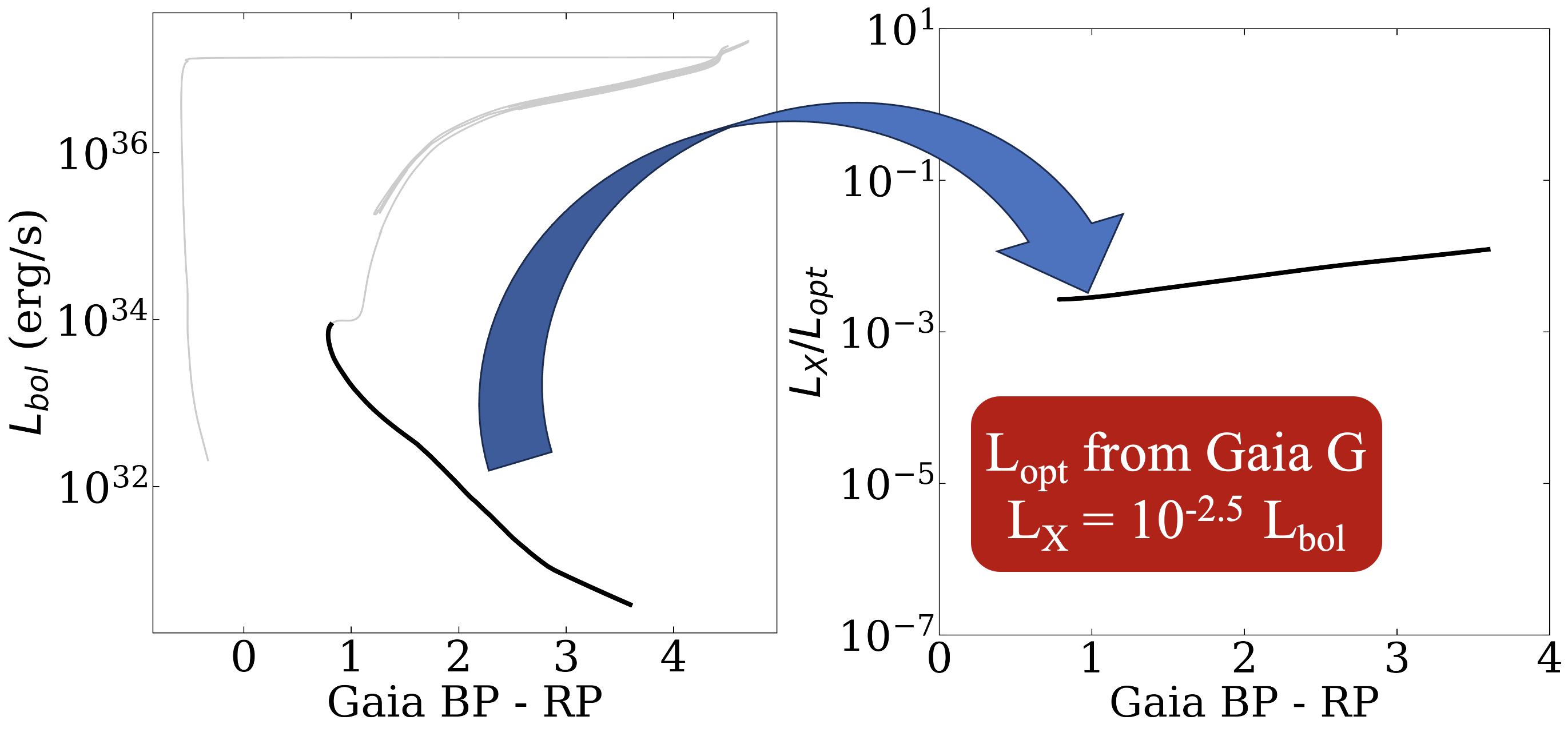}
    \caption{A MIST isochrone at the current age of the Milky Way converts between $L_\textrm{bol}$, an optical color, and $L_X$. The main sequence is shown in bold, and we omit the evolved tracks shown in lighter color.}
    \label{fig:hr}
\end{figure*}

Now that a limit of $L_X/L_\textrm{bol}$ is established, I convert $L_\textrm{bol}$ to a more useful observational quanity: an optical color and luminosity in a single optical bandpass, $L_\textrm{opt}$. To compute $L_\textrm{bol}$, I take a 10 Gyr isochrone from the MESA Isochrones and Stellar Tracks library \citep[MIST;][]{2016mist}, at solar metallicity and with $v/v_\textrm{crit}=0.4$. The final step is to obtain an optical luminosity, which can be taken at a single optical passband (i.e. \textit{Gaia} G). I do this by using the Sun's absolute G magnitude as a reference, and computing $L_\textrm{opt} = 10^{0.4 (M_{G, \odot} - M_G)}L_\odot$. The steps in creating the theoretical cut are visually outlined in Figure \ref{fig:hr}.

\subsection{Earlier Versions of This Diagram}
One of the earliest versions of this diagram appears in Figure 7 of \cite{1991stocke}. This diagram was created using data from the \textit{Einstein} X-ray observatory, and only a few dozen stellar sources appear in the plot. Nearly 25 years later, the diagram re-appeared in Figure 2 of \cite{2015greiner} using ROSAT data, now with a clear distinction between AGN and CVs in the upper left and stars in the lower right. However, just a few dozen sources are plotted, and only a limited explanation given as to why the diagram distinguishes between classes. Discouragingly, there appeared to be no way to distinguish between AGN and CVs, since they both occupied the same region of phase space (Gaia DR1 would come one year later). Perhaps the most up-to-date diagram is presented in Appendix B of \cite{2022stelzer}, where optical/X-ray counterparts from the eFEDS survey of the SRG/eROSITA telescope are shown. In that figure, extragalactic and Galactic objects are distinguished, but no further information is presented. 

I propose three main reasons for the lack of investigation and use of the diagram: 1) there have been fewer X-ray sources in previous surveys compared to today, 2) many of those sources did not even have reliable optical counterparts due to large X-ray error circles, and 3) distinguishing between Galactic and extragalactic sources was much more difficult before \textit{Gaia}. Many of the large catalogs on which I rely to provide classifications have also only come about in the last decade or so.

\subsection{Applications to Upcoming Large X-ray and Optical Surveys}

Given an X-ray detection and identification of an optical counterpart, this tool can efficiently select accreting compact objects candidates for optical spectroscopic follow-up. The efficiency is impressive: of the 25,050 sources in the XMM-Newton/\textit{Gaia} crossmatch, only 562 (2.2\%) of sources are above the empirical cut. Furthermore, because the vertical axis is written as $F_X/F_\textrm{opt}$, distant or faint objects that may only have a \textit{Gaia} proper motion (and not parallax) can still be selected.

\begin{figure}
    \centering
    \includegraphics[width=0.45\textwidth]{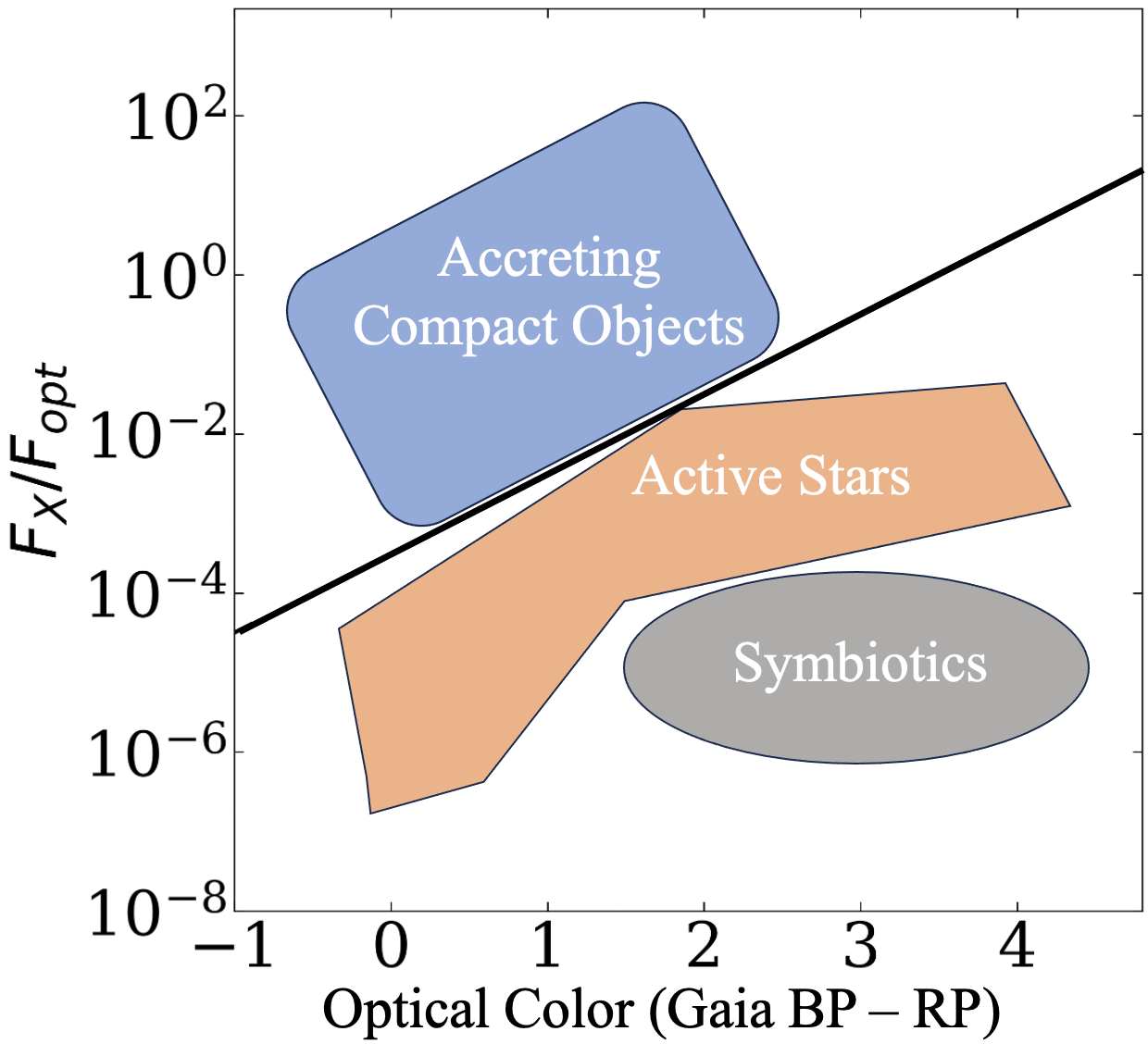}
    \caption{A cartoon of Figure \ref{fig:xms} demonstrating the location of the three main categories of objects in the X-ray Main Sequence.}
    \label{fig:cartoon}
\end{figure}

The most obvious application in the immediate future is for large scale spectroscopic surveys (e.g. SDSS-V, DESI, WEAVE, 4MOST). While these surveys are making obtaining optical spectra easier than ever, targeting is still needed to be done based on some prior information. I propose that these spectroscopic surveys inform their targeting strategies based on Figure \ref{fig:xms}. I present a cartoon outlining the main classes in Figure \ref{fig:cartoon}.

\section{Conclusion}

I have presented a tool for X-ray + optical astronomy that has received little to no attention in the past. The power of this tool is effectively demonstrated thanks to modern datasets --- in particular, 2" error circles from the large \textit{XMM-Newton} source catalog, precise astrometry from \textit{Gaia}, and a whole host of catalogs with object classifications in the literature. The ``X-ray Main Sequence" distinguishes accreting compact objects in the upper left from active stars and in the lower right. Symbiotic stars appear to occupy a unique portion of phase space as well.

I have presented two cuts to distinguish between accreting compact objects and active stars: an empirical cut and theoretical cut. The latter is based on a well-known relation in active stars, namely that they have a ``saturation limit" of $L_X/L_\textrm{bol} \lesssim 10^{-2.5}$ (though $L_X/L_\textrm{bol} \lesssim 10^{-3}$ is often quoted in the literature). The cuts I present are not perfect by any means, and could be affected by incompleteness in object catalogs that I use. 

In brief, the ``X-ray Main Sequence" shows promise for being used as an initial classification tool for upcoming surveys. It is highly efficient at selecting accreting compact objects, flagging a mere $\sim$2\% of objects as candidates. Both X-ray and optical astronomy are on the verge of transformation, thanks to SRG/eROSITA, \textit{Gaia}, the Rubin Observatory Legacy Survey of Space and Time (LSST), and the growing optical datasets from ZTF, TESS, ATLAS, and ASAS-SN. Stellar astronomy is entering a new golden era, and tools such as this will be crucial in identifying interesting single objects as well as large systematic surveys.

\section{Acknowledgements}
I wish to thank Shri Kulkarni and Kareem El-Badry for their thorough reading of this manuscript and useful feedback. I also thank Kevin Burdge, Adolfo Carvalho, Jim Fuller, Ilkham Galiullin, Lynne Hillenbrand, Rocio Kiman, Dovi Poznanski, Tom Prince, the ZTF Variable Star Group, and the members of various groups to whom I have presented this tool during its inception. I thank Zach Vanderbosch for a well-written tool for downloading SDSS spectra. I thank the staffs of the Palomar and Keck Observatories for their assistance in carrying out observations. I am grateful for support from an NSF Graduate Fellowship.

I thank the LSSTC Data Science Fellowship Program, which is funded by LSSTC, NSF Cybertraining Grant \#1829740, the Brinson Foundation, and the Moore Foundation; my participation in the program has benefited this work.

Based on observations obtained with XMM-Newton, an ESA science mission with instruments and contributions directly funded by ESA Member States and NASA. This work has made use of data from the European Space Agency (ESA) mission
{\it Gaia} (\url{https://www.cosmos.esa.int/gaia}), processed by the {\it Gaia}
Data Processing and Analysis Consortium (DPAC,
\url{https://www.cosmos.esa.int/web/gaia/dpac/consortium}). Funding for the DPAC
has been provided by national institutions, in particular the institutions
participating in the {\it Gaia} Multilateral Agreement.

Based on observations obtained with the Samuel Oschin Telescope 48-inch and the 60-inch Telescope at the Palomar Observatory as part of the Zwicky Transient Facility project. ZTF is supported by the National Science Foundation under Grants No. AST-1440341 and AST-2034437 and a collaboration including current partners Caltech, IPAC, the Weizmann Institute of Science, the Oskar Klein Center at Stockholm University, the University of Maryland, Deutsches Elektronen-Synchrotron and Humboldt University, the TANGO Consortium of Taiwan, the University of Wisconsin at Milwaukee, Trinity College Dublin, Lawrence Livermore National Laboratories, IN2P3, University of Warwick, Ruhr University Bochum, Northwestern University and former partners the University of Washington, Los Alamos National Laboratories, and Lawrence Berkeley National Laboratories. Operations are conducted by COO, IPAC, and UW. 

Funding for the Sloan Digital Sky Survey V has been provided by the Alfred P. Sloan Foundation, the Heising-Simons Foundation, the National Science Foundation, and the Participating Institutions. SDSS acknowledges support and resources from the Center for High-Performance Computing at the University of Utah. SDSS telescopes are located at Apache Point Observatory, funded by the Astrophysical Research Consortium and operated by New Mexico State University, and at Las Campanas Observatory, operated by the Carnegie Institution for Science. The SDSS web site is \url{www.sdss.org}.

SDSS is managed by the Astrophysical Research Consortium for the Participating Institutions of the SDSS Collaboration, including Caltech, The Carnegie Institution for Science, Chilean National Time Allocation Committee (CNTAC) ratified researchers, The Flatiron Institute, the Gotham Participation Group, Harvard University, Heidelberg University, The Johns Hopkins University, L’Ecole polytechnique f{\'e}d{\'e}rale de Lausanne (EPFL), Leibniz-Institut f{\"u}r Astrophysik Potsdam (AIP), Max-Planck-Institut f{\"u}r Astronomie (MPIA Heidelberg), Max-Planck-Institut f{\"u}r Extraterrestrische Physik (MPE), Nanjing University, National Astronomical Observatories of China (NAOC), New Mexico State University, The Ohio State University, Pennsylvania State University, Smithsonian Astrophysical Observatory, Space Telescope Science Institute (STScI), the Stellar Astrophysics Participation Group, Universidad Nacional Aut{\'o}noma de M{\'e}xico, University of Arizona, University of Colorado Boulder, University of Illinois at Urbana-Champaign, University of Toronto, University of Utah, University of Virginia, Yale University, and Yunnan University.

This work is based on data from eROSITA, the soft X-ray instrument aboard SRG, a joint Russian-German science mission supported by the Russian Space Agency (Roskosmos), in the interests of the Russian Academy of Sciences represented by its Space Research Institute (IKI), and the Deutsches Zentrum f{\"u}r Luft- und Raumfahrt (DLR). The SRG spacecraft was built by Lavochkin Association (NPOL) and its subcontractors, and is operated by NPOL with support from the Max Planck Institute for Extraterrestrial Physics (MPE). The development and construction of the eROSITA X-ray instrument was led by MPE, with contributions from the Dr. Karl Remeis Observatory Bamberg \& ECAP (FAU Erlangen-Nuernberg), the University of Hamburg Observatory, the Leibniz Institute for Astrophysics Potsdam (AIP), and the Institute for Astronomy and Astrophysics of the University of T{\"u}bingen, with the support of DLR and the Max Planck Society. The Argelander Institute for Astronomy of the University of Bonn and the Ludwig Maximilians Universit{\"a}t Munich also participated in the science preparation for eROSITA.

\bibliography{main}{}
\bibliographystyle{aasjournal}

\appendix

\section{New Cataclysmic Variables form XMM-Newton:Description of Observations and Individual Systems}
\label{sec:appendix}

All systems are summarized in Table \ref{tab:xmm_cvs}.

4XMM J001830.2+43571 (4XMMJ0018) was observed due to its 2.22 hr periodicity in ZTF. This system shows Balmer emission lines as well as prominent bumps centered at 4000 and 5000 Angstrom. These are reminiscent of cyclotron harmonics in polars, leading me to classify it as a candidate polar \citep[e.g.][]{1995warner}. The narrow emission lines, however, suggest this could be either a low-state polar or low accretion rate polar \citep[e.g.][]{2007schwope}.

4XMM J001912.5+220732 (4XMMJ0019) stood out due to its placement near the WD track in the \textit{Gaia} HR diagram. This object shows nearly no variability in ZTF, making it unlikely to have been discovered had it not been for its X-ray detection. It features strong, double-peaked Balmer and He emission lines, with weak H$\beta$ and H$\gamma$ absoprtion from the WD, leading me to classify it as a candidate WZ Sge CV \citep[e.g.][]{1986gilliland, 2023inight}. 

4XMM J021902.2+625713 (4XMMJ0219) stood out due to its placement near the main sequence despite having a 3.85 hr period in ZTF. The lack of strong Balmer emission lines (only weak H$\alpha$ is seen) as well as the FG-type spectrum is suggestive of the evolved CVs also known as pre-ELMs \citep{2021el-badry}. In these systems, the WD accretes from a donor that filled its Roche lobe just before leaving the main sequence, therefore forming an ``evolved" CV. Many systems evolve to short ($P_\textrm{orb}< 1$ hr) periods before detaching and forming an extremely low mass (ELM) WD \citep{2021el-badry}. Virtually all of the criteria are met for me to classify this system as an (pre-ELM) evolved CV. 

4XMM J063722.6+054158 (4XMMJ0637)  was selected due to its long orbital period ($P_\textrm{orb}= 13.8$ hr) dominated by ellipsoidal modulation. It also undergoes regular, triangular outbursts, which last $\sim 200$ days, in contrast to typical dwarf nova outbursts which last $\sim 10$ days. At its orbital period, the donor must be an evolved subgiant \citep[e.g.][]{2022evolved}. The optical spectrum is indeed dominated by the FG-type donor, albeit with He II 4686 and the rare CIII/NIII Bowen blend emission lines. Interestingly, this system is near the ``bifurcation period", where depending on the WD and donor parameters, the system may either evolve to longer or shorter orbtial periods \citep[e.g.][]{2023belloni}. All of the criteria are met for me to classify this system as an evolved CV, and I will report on extensive spectroscopy in an upcoming study (Rodriguez et al. in prep).

4XMM J085012.5-03163 (4XMMJ0850) stood out due to its 51 min (or doubled, 1.70 hr) period in ZTF data. Phase-resolved spectroscopy is being acquired to determine the true period. Like 4XMMJ0219, this system is also unusually close to the main sequence for so short an orbital period (even if 1.70 hr is the true period). The spectrum is dominated by a K-type donor, which hints at it being an evolved CV. In this system, however, Balmer lines are in emission along with the He II 4686 line, which is not typically seen in pre-ELMs, but suggestive of magnetism \citep{1992silber, 2021el-badry}. While some characteristics lead me to classify this system as a candidate evolved CV, further work is needed to determine its true nature.

4XMM J195502.7+443657 (4XMMJ1955) features unusual variability in ZTF --- low amplitude ($< 1$ mag) variability with high amplitude ($\sim 3$ mag), short ($\sim 20$ day) dips. High-cadence ZTF data also reveals a 20--30 min period, suggestive of a WD spin period in intermediate polars \citep{2017mukai}. Remarkably, its optical spectrum shows the strongest Balmer and He emission lines of any object in the sample. He II 4686 is also nearly as strong as H$\beta$, which is typically seen in intermediate polars \citep{1992silber}. Despite fulfilling some of the necessary criteria, I am cautious to label this system as a candidate intermediate polar since extensive X-ray and optical timing data as well as phase-resolved spectroscopy as usually needed to securely classify such systems.

\begin{figure}
    \centering    \includegraphics[width=0.48\textwidth]{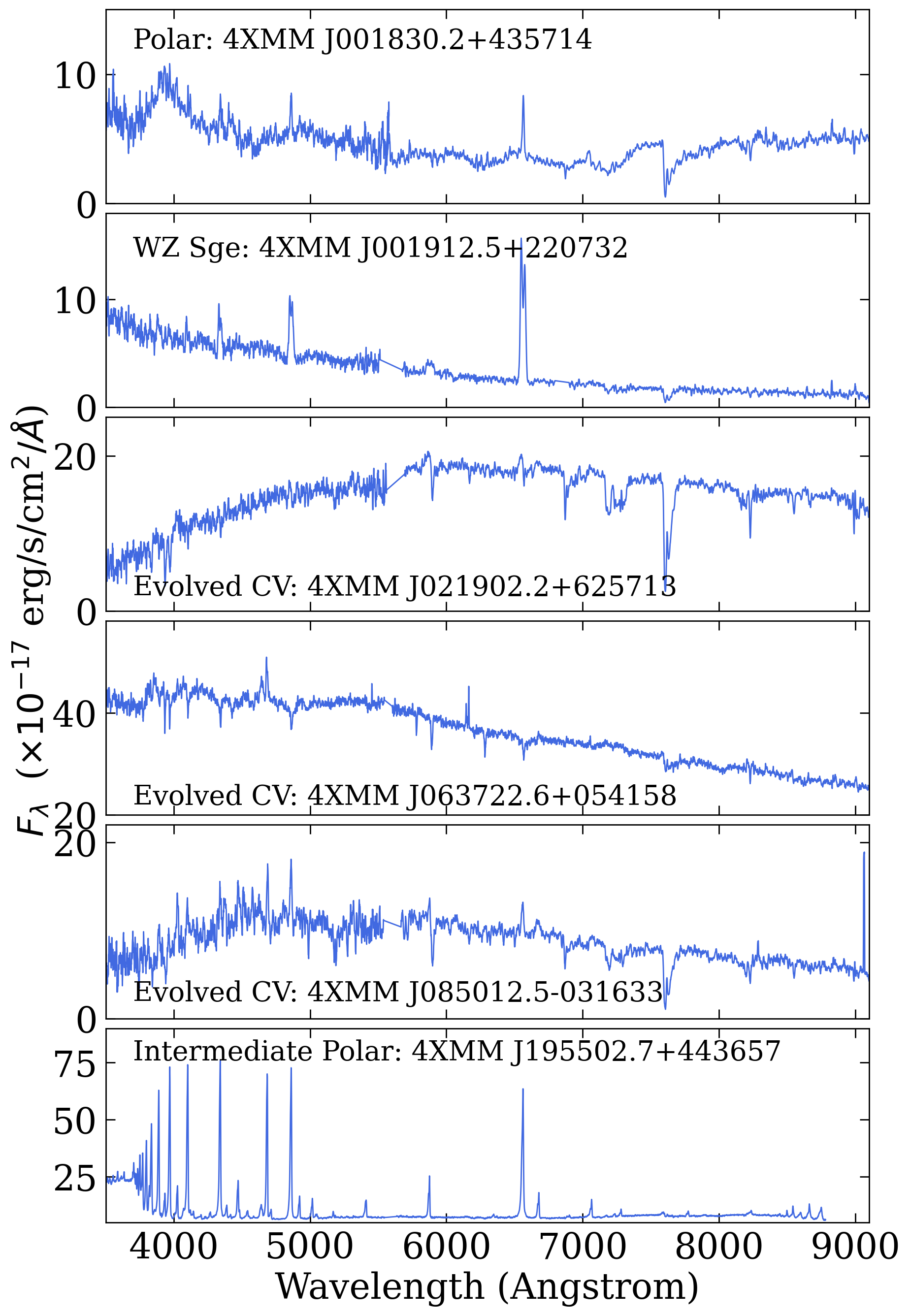}
    \caption{Keck and Palomar spectra of six CVs which are part of an ongoing spectroscopic survey of the 4XMM-\textit{Gaia} catalog. All sources were predicted to be accreting compact objects using the X-ray Main Sequence, and confirmed through optical spectroscopy. Sub-classes shown are preliminary, yet all systems are distinct from archetypal CVs (dwarf novae such as that shown in Figure \ref{fig:efeds_sdss_spectra}), demonstrating the richness that X-ray + optical surveys can reveal.}
    \label{fig:xmm_spectra}
\end{figure}

Of the six systems reported above, four were observed with the Double Spectrograph \citep[DBSP;][]{1982dbsp} on the Hale telescope (4XMMJ0018, 4XMMJ0019 4XMMJ0219, and 4XMMJ0850). I used the 600/4000 grism on the blue side and the 316/7500 grating on the red side. A 1.5" slit was used, and the seeing throughout all observations varied between 1.5 -- 2.0", leading to some slit losses. All P200/DBSP data were reduced with \texttt{DBSP-DRP}\footnote{\url{https://dbsp-drp.readthedocs.io/en/stable/index.html}}, a Python-based pipeline optimized for DBSP built on the more general \texttt{PypeIt} pipeline \citep{2020pypeit}. 

4XMMJ0637 and 4XMMJ1955 were observed with the Keck I telescope using the Low-Resolution Imaging Spectrometer \citep[LRIS;][]{1995lris}. I used the 600/4000 grism on the blue side with 2x2 binning (spatial, spectral), and the 600/7500 grating on the red side with 2x1 binning. I used a 1.0$\arcsec$ slit, and the seeing each night was approximately 0.7--1$\arcsec$, leading to minimal slit losses. All Keck I/LRIS data were reduced with \texttt{lpipe}, an IDL-based pipeline optimized for LRIS long slit spectroscopy and imaging \citep{2019perley_lpipe}. All data (for both DBSP and LRIS) were flat fielded sky-subtracted using standard techniques. Internal arc lamps were used for the wavelength calibration and a standard star for overall flux calibration. 

\begin{figure*}
    \centering
    \includegraphics[width=0.9\textwidth]{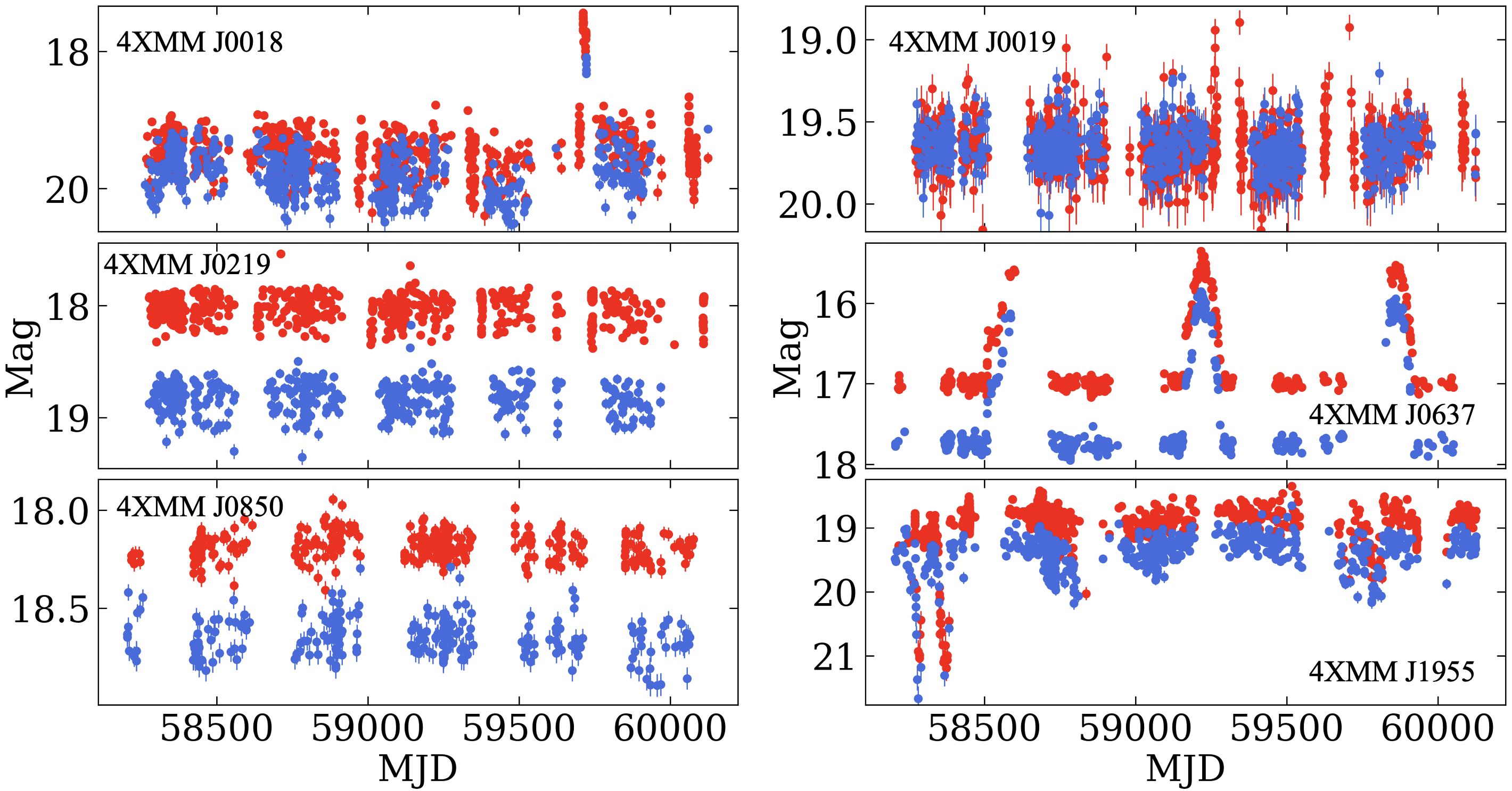}
    \caption{Long term (5 year) ZTF light curves in $r$ band (red) and $g$ band (blue) of all newly discovered XMM CVs.}
    \label{fig:xmm_lc}
\end{figure*}

In Figure \ref{fig:xmm_lc}, I present long term ZTF light curves for all CVs in Section \ref{sec:xmm_cv}, from Data Release 19 (covering March 2018 -- July 2023). ZTF is a photometric survey that uses a wide 47 $\textrm{deg}^2$ field-of-view camera mounted on the Samuel Oschin 48-inch telescope at Palomar Observatory with $g$, $r$, and $i$ filters \citep{bellm2019, graham2019, dekanyztf, masci_ztf}. In its first year of operations, ZTF carried out a public nightly Galactic Plane Survey in $g$-band and $r$-band \citep{ztf_northernskysurvey_bellm, kupfer_ztf}. This survey was in addition to the Northern Sky Survey which operated on a 3 day cadence \citep{bellm2019}. Since entering Phase II, the public Northern Sky Survey is now at a 2-day cadence. The pixel size of the ZTF camera is 1$\arcsec$ and the median delivered image quality is 2.0$\arcsec$ at FWHM. 

\begin{figure*}
    \centering
    \includegraphics[width=0.9\textwidth]{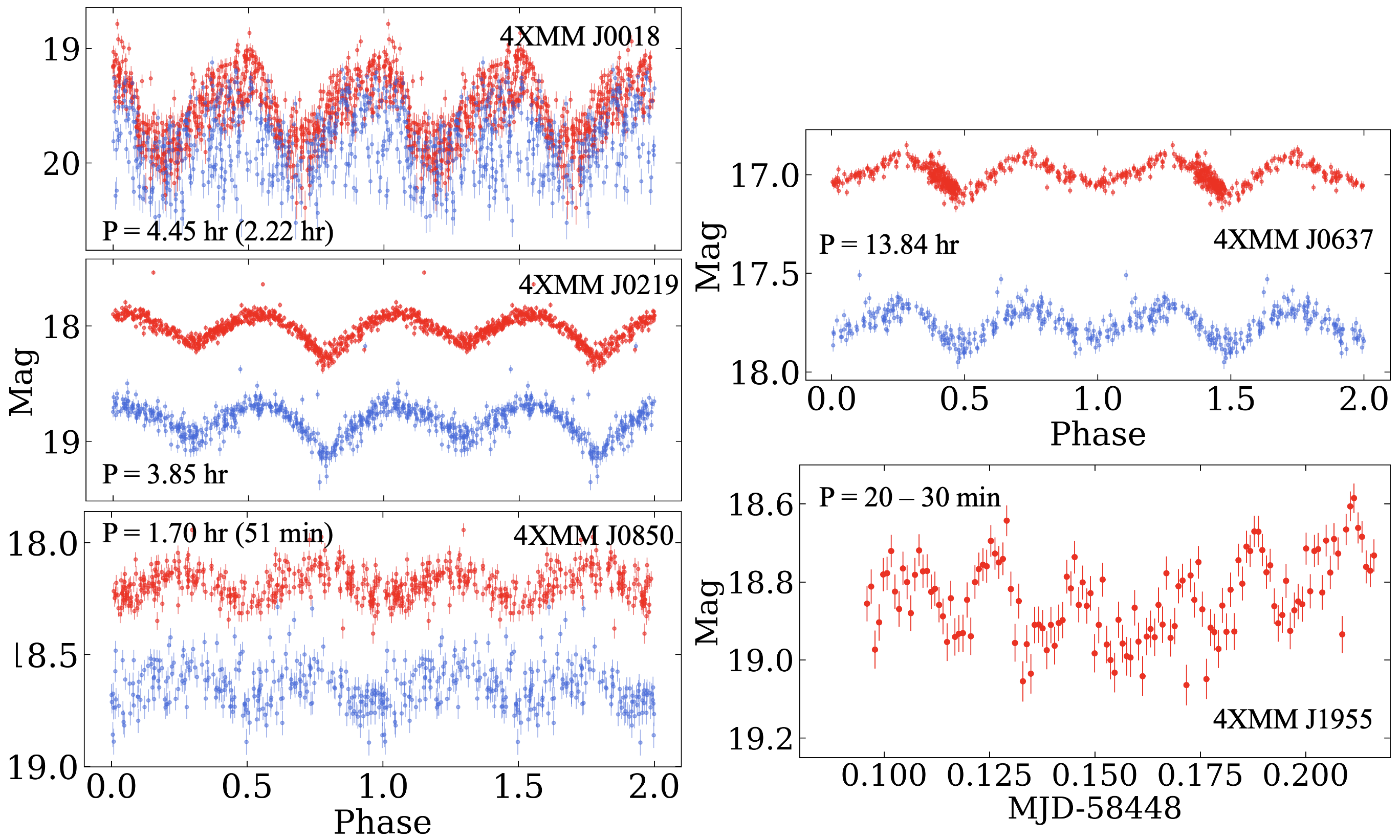}
    \caption{Folded ZTF light curves (excluding outbursts) in $r$ band (red) and $g$ band (blue) of 4 newly discovered XMM CVs. On the bottom right are the high-cadence continuous observations of 4XMMJ1955 from the ZTF Galactic Plane Survey.}
    \label{fig:xmm_phot}
\end{figure*}

In Figure \ref{fig:xmm_phot}, I present phase-folded light curves for four objects: 4XMMJ0018, 4XMMJ0219, 4XMMJ0850, and 4XMMJ0637. I used the \texttt{gatspy} software \citep{2015gatspy} to compute a Lomb-Scargle periodogram \citep{1982scargle}, searching for periods between 4 minutes and 10 days with an oversampling factor of 5. For all four systems that show a significant ($10\sigma$ above the median) periodicity, I plot twice the best period, which in 4XMMJ0219 and 4XMMJ0637 reveals ellipsoidal modulations (minima of different depths). However, in the case of 4XMMJ0018 and 4XMMJ0850, the difference between two different minima is unclear and phase-resolved spectroscopy is needed to reveal the true period. Finally, I present the high-cadence data from the ZTF Galactic Plane Survey for 4XMMJ1955, which reveals pulsations on a 20--30 min timescale.

\begin{table}[]
    \centering
\begin{tabular}{lrrll}

                 Name &    Gaia RA (deg) &  Gaia DEC (deg) &          Classification & Orbital Period (hr) \\
                 \hline
4XMM J001830.2+435714 &   4.626476 & 43.954572 &              CV (Polar) &         2.22 (4.45) \\
4XMM J001912.5+220732 &   4.802370 & 22.125808 &             CV (WZ Sge) &                 - \\
4XMM J021902.2+625713 &  34.760214 & 62.953643 &            CV (evolved) &                3.85 \\
4XMM J063722.6+054158 &  99.344615 &  5.699567 &            CV (evolved) &                13.8 \\
4XMM J085012.5-031633 & 132.552197 & -3.276101 &            CV (evolved) &         1.70 (0.85) \\
4XMM J195502.7+443657 & 298.761724 & 44.616101 & CV (Intermediate Polar) &                 - \\

\end{tabular}
    \caption{Summary of all newly discovered XMM CVs. The most likely orbital period is listed first, and the one in parentheses cannot be ruled out without additional data.}
    \label{tab:xmm_cvs}
\end{table}
\newpage
\section{Catalog Data}

 In Table \ref{tab:other_objects}, I list all of the objects used from external catalogs and references to the papers in which X-ray fluxes were reported. In Table \ref{tab:xmm_gaia}, I show a preview of the entire XMM-\textit{Gaia} crossmatched catalog, which I make available in machine readable format.

\begin{table}[]
    \centering
\begin{tabular}{llrl}
                   Name &          Class &             Gaia DR3 ID & X-ray Reference \\
\hline
             J0212+5320 &        Redback &  455282205716288384 &             (1) \\
             J1048+2339 &        Redback & 3990037124929068032 &             (1) \\
               J1306-40 &        Redback & 6140785016794586752 &             (1) \\
             J1431-4715 &        Redback & 6098156298150016768 &             (1) \\
             J1622-0315 &        Redback & 4358428942492430336 &             (1) \\
             J1628-3205 &        Redback & 6025344817107454464 &             (1) \\
             J1723-2837 &        Redback & 4059795674516044800 &             (1) \\
             J1803-6707 &        Redback & 6436867623955512064 &             (1) \\
             J1816+4510 &        Redback & 2115337192179377792 &             (1) \\
             J1908+2105 &        Redback & 4519819661567533696 &             (1) \\
             J1910-5320 &        Redback & 6644467032871428992 &             (1) \\
             J2039-5618 &        Redback & 6469722508861870080 &             (1) \\
             J2129-0429 &        Redback & 2672030065446134656 &             (1) \\
             J2215+5135 &        Redback & 2001168543319218048 &             (1) \\
             J2339-0533 &        Redback & 2440660623886405504 &             (1) \\
             J1311-3430 &    Black Widow & 6179115508262195200 &             (1) \\
             J1653-0158 &    Black Widow & 4379227476242700928 &             (1) \\
             J1810+1744 &    Black Widow & 4526229058440076288 &             (1) \\
               B1957+20 &    Black Widow & 1823773960079216896 &             (1) \\
            GROJ0422+32 &      LMXB (BH) &  172650748928103552 &             (2) \\
               A0620-00 &      LMXB (BH) & 3118721026600835328 &             (2) \\
               V404 Cyg &      LMXB (BH) & 2056188624872569088 &             (2) \\
          XTE J1118+480 &      LMXB (BH) &  789430249033567744 &             (2) \\
            GROJ1655-40 &      LMXB (BH) & 5969790961312131456 &             (2) \\
             4U 2129+47 &      LMXB (NS) & 1978241050130301312 &             (3) \\
                Cen X-4 &      LMXB (NS) & 6205715168442046592 &             (3) \\
                Aql X-1 &      LMXB (NS) & 4264296556603631872 &             (3) \\
       SAX J1808.4-3658 &      LMXB (NS) & 4037867740522984832 &             (3) \\
               A0535+26 &      HMXB (NS) & 3441207615229815040 &             (4) \\
            KS 1947+300 &      HMXB (NS) & 2031939548802102656 &             (4) \\
              V4641 Sgr &      HMXB (BH) & 4053096388919082368 &             (4) \\
                Cyg X-1 &      HMXB (BH) & 2059383668236814720 &             (4) \\
                 GX 1+4 & Symbiotic (NS) & 4110236324513030656 &             (5) \\
            4U 1954+319 & Symbiotic (NS) & 2034031438383765760 &             (5) \\
CXOGBS J173620.2–293338 & Symbiotic (NS) & 4060066227422719872 &             (5) \\
             4U 1700+24 & Symbiotic (NS) & 4571810378118789760 &             (5) \\
                 NQ Gem & Symbiotic (WD) &  868424696282795392 &             (6) \\
                 UV Aur & Symbiotic (WD) &  180919213811383680 &             (6) \\
                 ZZ CMi & Symbiotic (WD) & 3155368612444708096 &             (6) \\
                 ER Del & Symbiotic (WD) & 1750795043999682304 &             (6) \\
             CD -283719 & Symbiotic (WD) & 5608089951177429120 &             (6) \\
        RX J0019.8+2156 &            SSS & 2800287654443977344 &             (7) \\
        RX J0925.7-4758 &            SSS & 5422337322910734080 &             (7) \\
                 RR Tel &            SSS & 6448785024330499456 &             (7) \\
\end{tabular}
    \caption{All systems shown in Figure \ref{fig:xms} with literature X-ray detections. (1): \cite{2023spiders}, (2): \cite{2001garcia}, (3): \cite{1999menou}, (4): \cite{2006russell}, (5): \cite{2019yungelson}, (6): \cite{2013luna}, (7): 4XMM-DR13 Catalog.}
    \label{tab:other_objects}
\end{table}

\begin{table}[]
    \centering
\begin{tabular}{rlrlll}

   4XMM-DR13 ID &               IAUNAME &         Gaia DR3 ID &  Non-magnetic CV &   YSO &  Active Binary \dots \\
\hline
201253101010037 & 4XMM J000009.8-251920 & 2335010480014243328 &            False & False &          False \\
204033901010033 & 4XMM J000012.9+622946 &  429950007577101952 &            False & False &          False \\
207009901010018 & 4XMM J000014.6+675337 &  528609770043800832 &            False & False &          False \\
206584004010005 & 4XMM J000024.9+443634 &  385045506010558336 &            False & False &          False \\
207009901010019 & 4XMM J000030.3+681159 &  528996042222710016 &            False & False &          False \\
207009901010001 & 4XMM J000032.0+681500 &  528996591978497408 &            False & False &          False \\
206935404010007 & 4XMM J000047.9+233216 & 2848395067729986688 &            False & False &          False \\
206584004010038 & 4XMM J000102.0+442809 &  385041279762788608 &            False & False &          False \\
206584004010048 & 4XMM J000110.9+443700 &  385043405769547136 &            False & False &          False \\
204033901010077 & 4XMM J000129.2+622456 &  429945643890463488 &            False & False &          False \\
200161401010017 & 4XMM J000130.0+625236 &  430053396031738112 &            False & False &          False \\
206584004010017 & 4XMM J000139.7+442610 &  385028910256998528 &            False & False &           True \\
206935404010018 & 4XMM J000146.9+233512 & 2848232133851323264 &            False & False &          False \\
200417501010032 & 4XMM J000153.4-300612 & 2320841520343862400 &            False & False &          False \\
204033901010016 & 4XMM J000153.6+623220 &  429947499316231808 &            False & False &          False \\
200417501010078 & 4XMM J000154.4-301038 & 2320834614036455424 &            False & False &          False \\
200417501010007 & 4XMM J000154.4-300741 & 2320835576109126400 &            False & False &          False \\
204033901010056 & 4XMM J000155.6+622800 &  429946812121535488 &            False & False &          False \\
200417501010112 & 4XMM J000157.0-301209 & 2320834304798813568 &            False & False &          False \\
200417501010138 & 4XMM J000157.6-300929 & 2320834751475799808 &            False & False &          False \\
\dots &&&&&

\end{tabular}
    \caption{I make the full XMM-\textit{Gaia} crossmatch freely available in machine readable format. The first twenty rows, and select columns are shown here as a preview.}
    \label{tab:xmm_gaia}
\end{table}

\end{document}